# Erdmessung mit Quanten und Relativität

Prof. Dr.-Ing. habil. Jürgen Müller
Institut für Erdmessung, Leibniz Universität Hannover
Schneiderberg 50, 30167 Hannover
E-Mail: mueller@ife.uni-hannover.de

## 1. Einführung

In der Physikalischen Geodäsie wird die Erde als Ganzes vermessen und die gewonnen Daten analysiert. Dabei werden verschiedene Messverfahren eingesetzt auf der Erde und im Weltraum, und es werden unterschiedliche Phänomene erfasst, wie etwa die Figur der Erde, Schwankungen der Erdrotation oder das Schwerefeld. Der Begriff Erdmessung fasst dies recht kompakt zusammen.

Zum einen wird die Form der Erde geometrisch erfasst, also ein Abbild der Oberfläche und damit ihre Figur abgeleitet, wobei man die höchste Genauigkeit für die Standorte der so genannten geodätischen Weltraumverfahren (z.B. GNSS, VLBI, SLR) erhält. Hier werden aus Beobachtungen von Satelliten oder stellaren Objekten Koordinaten der Observatorien mit mm-Genauigkeit bestimmt. Ein bekanntes Verfahren ist GPS (Global Positioning System). Es gehört zur Gruppe der GNSS (Global Navigation Satellite Systems), deren Mikrowellen-Signale über eine Phasendifferenzmessung ausgewertet werden. GNSS werden zur weltweiten Positionierung und Navigation eingesetzt. Bei VLBI (Very Long Baseline Interferometry) wird eine ganz andere Methode angewandt: Die Interferometrie auf langen Basislinien. Die Signale ferner Radioquellen werden von mehreren Teleskopen auf der Erde empfangen. Durch die Auswertung der Differenzen der Ankunftszeiten der Signale können einerseits Erdorientierungsparameter und Stationskoordinaten im globalen terrestrischen Referenzsystem mit mm-Genauigkeit bestimmt werden und andererseits die Positionen der Radioquellen als raumfeste Referenz. Bei SLR (Satellite Laser Ranging) werden dagegen die Laufzeiten von Laserpulsen zu Satelliten und zurück mm-genau gemessen, um daraus Stationskoordinaten sowie Satellitenbahnen und Schwerefeldparameter zu bestimmen.

Da man immer Objekte außerhalb der Erde (Satelliten oder Radioquellen) als äußere Referenz verwendet, lassen sich Bewegung und Rotationsschwankungen der Erde sehr gut beobachten. Aus der Analyse der Daten der geodätischen Weltraumverfahren können somit als primäre Zielparameter, neben den zuvor genannten Stationskoordinaten und ihren Geschwindigkeiten, Erdorientierungsgrößen, wie Präzession/Nutation sowie Polbewegung und Tageslängenschwankungen, abgeleitet und zugrunde liegende dynamische Prozesse untersucht werden (Seitz und Müller 2016).

Neben der geometrischen Betrachtung kann die Figur der Erde auch physikalisch definiert und bestimmt werden. Hier wird dann das räumlich und zeitlich variable

Schwerefeld der Erde beobachtet, um etwa Äquipotentialflächen, z.B. das Geoid, das mit der mittleren Meeresoberfläche in Ruhe zusammenfällt, zu bestimmen. Für die langwelligen Strukturen des Schwerefeldes eignet sich wieder SLR. Für eine höhere räumliche Auflösung, also die Erfassung kleinerer Strukturen des Schwerefelds, sowie zeitliche Änderungen wurden seit 2000 spezielle Satellitenmissionen realisiert; siehe auch Torge und Müller (2012) sowie Pail et al. (2015) und dort angegebene Referenzen. Die ESA-Mission GOCE (Gravity field and steady-state Ocean Circulation Explorer) lieferte zwischen 2009 und 2013 Daten, die es erlaubten das Geoid mit einer Genauigkeit von 1 – 2 cm für räumliche Strukturen von 100 km auf der Erdoberfläche zu berechnen. Als primäres Messkonzept wurde die Gradiometrie realisiert; hier werden differentielle Gravitationsbeschleunigungen im Satelliten mit höchster Präzision beobachtet (Rummel et al. 2011). GOCE liefert damit die Grundlage zur Errichtung globaler, physikalisch definierter Höhensysteme sowie – zusammen mit der altimetrischen Bestimmung des Meeresspiegels – zur genauen Ableitung von Meeresströmungen. Die Mission GRACE (Gravity Recovery and Climate Experiment), geleitet von NASA und DLR, ist seit 2002 im Orbit. Messmethode ist das Satellite-to-Satellite-Tracking, wobei die variierenden Abstände zwischen den beiden Satelliten mit Hilfe eines Mikrowellensystems mit µm-Genauigkeit beobachtet werden. GRACE liefert monatliche Schwerefelder und erlaubt somit, zeitliche Variationen zu analysieren, deren Ursache etwa im Abschmelzen der Eismassen der großen kontinentalen Schilde (Grönland, Antarktis) liegt; allein in Grönland beträgt der Eismassenverlust 200 bis 300 Gigatonnen pro Jahr. Ein anderes Anwendungsbeispiel ist die Erfassung von Veränderungen im hydrologischen Wasserkreislauf, z.B. durch übermäßige Grundwasserentnahme im Iran oder Nordindien für Bewässerungszwecke; exemplarisch sei Eicker et al. (2016) genannt.

Da bei den bisherigen Satellitenmissionen weder die räumliche noch zeitliche Auflösung ausreicht, um die Anforderungen in den diversen Geo-Disziplinen zu erfüllen und die komplexen, interagierenden Prozesse zu verstehen (Pail et al. 2015), werden bereits Konzepte für Nachfolgemissionen entwickelt und teilweise umgesetzt. Ziel ist es, eine höhere Auflösung, eine größere Genauigkeit und generell längere Zeitreihen zu erhalten (NGGM-D 2014, Elsaka et al. 2014). Bei einer dieser Missionen, GRACE-FO, soll der Abstand zwischen den beiden Satelliten mittels Laserinterferometrie nanometer-genau gemessen werden (Sheard et al. 2012, Abb. 1), wobei auch Technologie aus Hannover zum Einsatz kommt, die ursprünglich im Hinblick auf die Gravitationswellendetektion entwickelt wurde.

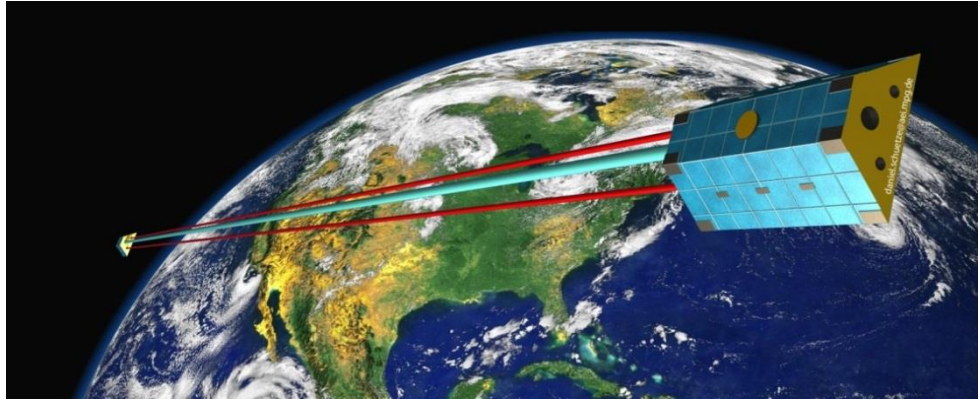

Abb. 1: Künstlerische Darstellung der GRACE-FO-Mission; illustriert sind die Mikrowellenabstandsmessung zwischen den beiden Satelliten, wie sie auch bei GRACE realisiert wurde (der mittlere Strahl), sowie die neuartige Laserinterferometrie (die beiden außen verlaufenden Strahlen).

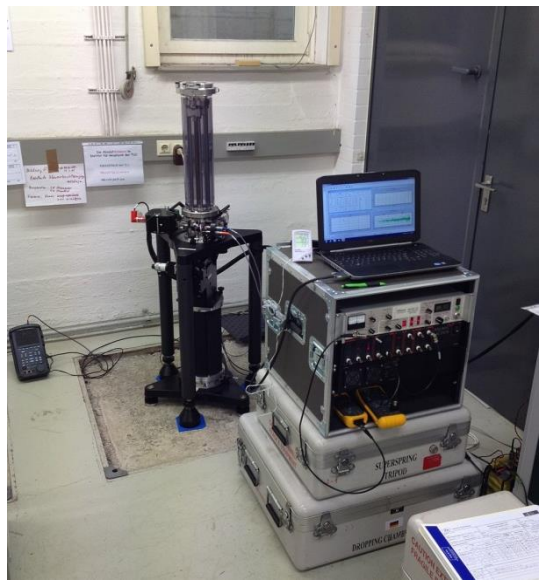

Abb. 2: Absolutgravimeter FG5X-220 des Instituts für Erdmessung bei einer Messung an der TU Clausthal, Institut für Geophysik

Die ganz feinen Strukturen im Schwerefeld kann man nicht vom Weltraum aus erfassen, sie werden durch terrestrische Messungen ergänzt. Hier kommen Gravimeter, wie das FG5X von Micro-g LaCoste (Abb. 2), zum Einsatz. Über gleichzeitige Zeit- und Abstandmessung eines in einer Vakuumröhre frei fallenden Reflektors wird mit Laserinterferometrie der lokale Schwerewert $g$ bestimmt. Man erreicht Genauigkeiten von ca. 2 µgal (= 20 nm s$^{-2}$), wobei man dafür etwa 2 Tage beobachten muss, um systematische Fehler, z.B. aufgrund seismischer Störungen, zu reduzieren. Aus Wiederholungsmessungen lassen sich für den Aufstellungsort zeitabhängige Massenvariationen bestimmen. Ein Beispiel ist die fennoskandische Landhebung, die über mehrjährige Messkampagnen (Gitlein 2009) erfasst wurde. Vergleiche mit GRACE-Analysen (Steffen et al. 2009, Timmen et al. 2011) zeigen noch Differenzen (Abb. 3), speziell in den Randbereichen, in denen die Signale geringer sind. Um das Phänomen als Ganzes zu verstehen, sind die diversen Datentypen zu kombinieren (Müller et al. 2012). Die skandinavischen Kollegen haben auch geometrische GNSS-Beobachtungen und Nivellements in die Lösung für ihr neues Landhebungsmodell NKG2016LU integriert (Vestøl et al. 2016).

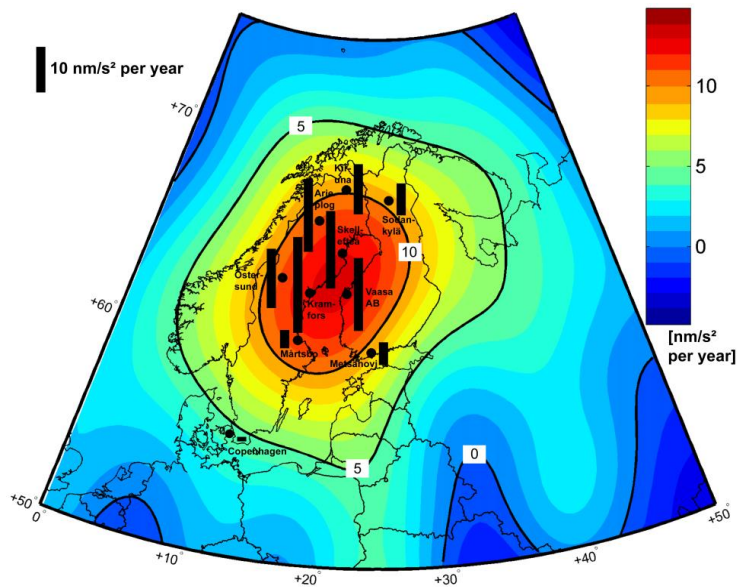

Abb. 3: Nacheiszeitlicher Ausgleichsprozess in Skandinavien; farbig hinterlegt sind die aus GRACE abgeleiteten Massenvariationen; die schwarzen Balken geben die in einzelnen Punkten mittels terrestrischer Absolutgravimetrie gemessenen Werte an (Gitlein 2009).

Um nun noch höhere Genauigkeiten oder eine bessere Auflösung zu erreichen, wurde in den letzten Jahren an ganz neuen Messkonzepten auf der Erde und im Weltraum gearbeitet. Die Nutzung der Laserinterferometrie im Weltraum zur präzisen Abstandmessung zwischen zwei Satelliten wurde bereits erwähnt (siehe z.B. auch Reubelt et al. 2014). Der Kern einer weiteren Methode ist die Atominterferometrie, die zur Schweremessung genutzt werden kann, indem man frei fallende Atome beobachtet. Außerdem können mit Hilfe präziser Atomuhren unter Ausnutzung der Relativitätstheorie Potentialunterschiede im Schwerefeld gemessen werden. Die beiden zuletzt genannten innovativen Methoden werden im Folgenden näher erläutert.

## 2. Atominterferometrische Gravimetrie

Bei einem Quantengravimeter lässt man nicht makroskopische Objekte wie Retroreflektoren fallen, sondern Atome oder Atomwolken; für eine relativ einfache Beschreibung siehe Schilling et al. (2012). Die Atome werden zunächst stark gekühlt, um sie besser kontrollieren zu können (Abb. 4).

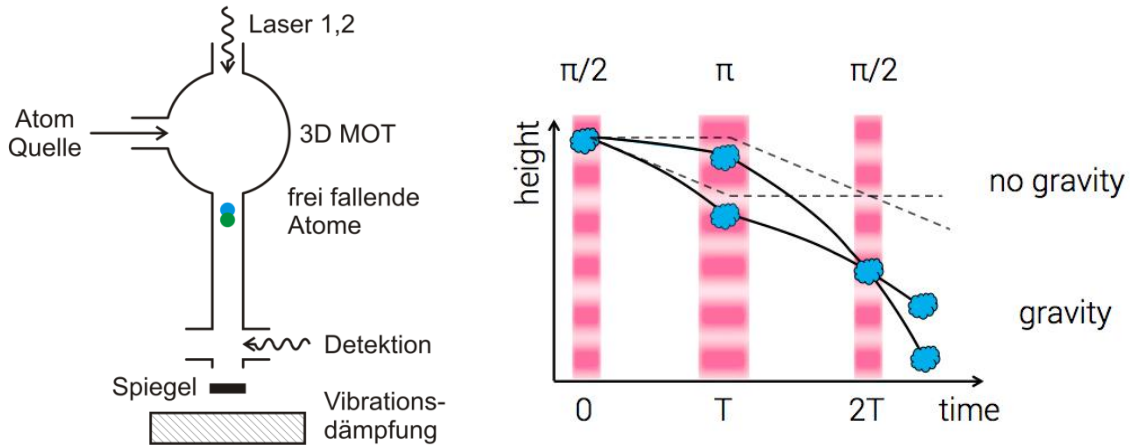

Abb. 4: Links: Schematischer Aufbau eines Atomgravimeters (aus Schilling et al. 2012), rechts: Schema der Atominterferometrie für die Bestimmung der Schwerebeschleunigung g (Rasel 2014, priv. Komm.)

Die Atome werden in einer magneto-optischen Falle ('MOT') gefangen und auf wenige µKelvin gekühlt. Nach Abschalten der Falle beginnt der freie Fall in einer evakuierten Fallkammer. Die Atome werden mit einem Puls zweier gegenläufiger Laserstrahlen manipuliert, was zum Aufteilen der Atome führt (etwa die Hälfte der Atome laufen dann statistisch auf einem anderen Weg); der Laserimpuls fungiert dabei wie ein Strahlteiler bei der klassischen Interferometrie. Nach einer gewissen Zeit *T* werden die Atome wieder mit einem Laserpuls ‚abgelenkt' (der Spiegel des Interferometers), so dass sie sich nach der Zeit 2*T* treffen und interferieren (der abschließende Strahlteiler). Die beobachtete Phasenverschiebung $\Delta\varphi$ ist dann direkt proportional der lokalen Schwerebeschleunigung **g** ($k_{eff}$ ist der effektive Wellenvektor des Laserlichtes, der senkrecht auf der Wellenfront steht):

$$\Delta\varphi = -\boldsymbol{k}_{eff}\boldsymbol{g}T^2. \tag{1}$$

Der Vorteil einer atominterferometrischen Schweremessung besteht darin, dass manche systematische Effekte geringer sind, da man keine großen fallenden Objekte hat, die zu Störungen führen. Man kann einen Schwerewert *g* = |**g**| innerhalb weniger Minuten mit hoher Genauigkeit von 1 µgal erhalten. Die schnellere Vermessung ermöglicht neue Beobachtungskonzepte. Aktuelle Entwicklungen laufen an der Leibniz Universität Hannover sowie an der Humboldt Universität Berlin (Hauth et al. 2013, Rudolph et al. 2015). Die französische Firma µquans verkauft sogar schon robuste Quantengravimeter (www.muquans.com). Das Berliner Instrument GAIN arbeitet nach dem Wurf-Prinzip. Vergleichsmessungen mit klassischen Gravimetern, etwa 2015 in Onsala (Freier et al. 2016), zeigen das hohe Potential atominterferometrischer Gravimeter deutlich (Abb. 5). Man kann sie sowohl als Absolutgravimeter einsetzen, die einen absoluten lokalen Schwerewert *g* in sehr kurzer Zeit liefern, wie auch als registrierende Gravimeter, um über einen längeren Zeitraum Schwerevariationen zu erfassen. Somit verknüpft dieser neuartige Gravimetertyp die Vorteile von Absolut- und supraleitenden Gravimetern in einem Gerät. Dieses Messkonzept könnte außerdem aufgrund seiner Qualitäten künftig als neuer Standard für Schweremessungen dienen und würde auch die präzise Bestimmung von Kalibrierparametern für supraleitende Gravimeter mit besserer Genauigkeit erlauben.

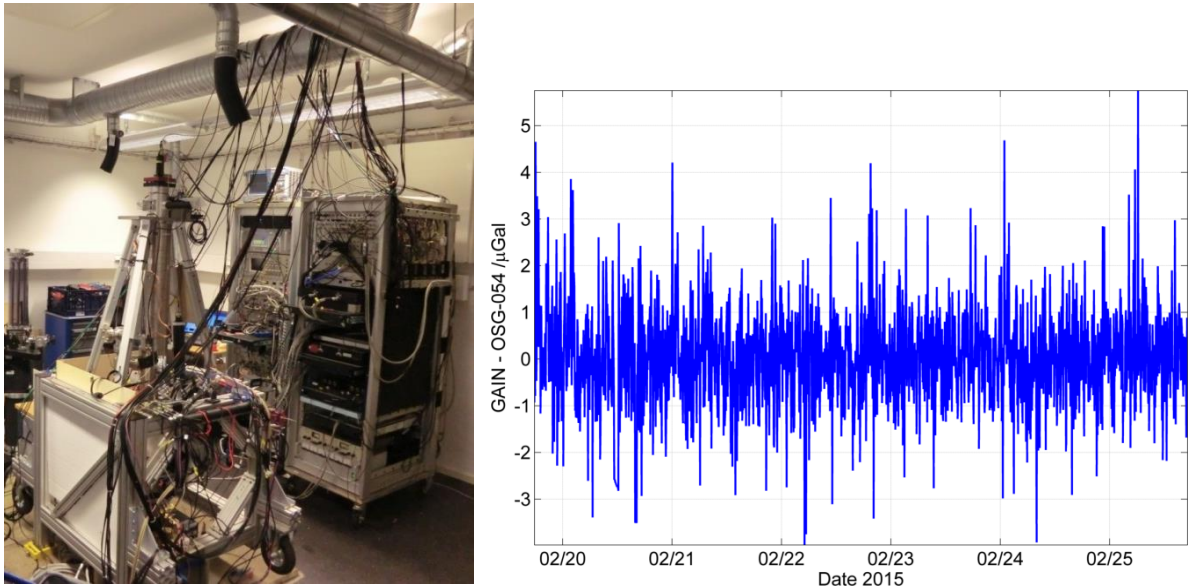

Abb. 5: Links: Das Quantengravimeter GAIN während der Messung in Onsala (links im Hintergrund das FG5X-220); rechts: Vergleichsmessung zu einem Supraleitgravimeter, die Differenz entspricht einer Messgenauigkeit der Schwere *g* von 1 µgal über 5 Minuten (Schilling 2015, priv. Komm.)

Diese Messtechnologie basierend auf Atominterferometrie liefert nicht nur Schwerebeschleunigungen, sondern kann in einem entsprechenden Aufbau auch kinematische Beschleunigungen, inklusive Rotationsbeschleunigungen erfassen, was wiederum interessante Anwendungen in der Navigation eröffnet. Der große Vorteil hier ist, dass die Atominterferometrie im idealen Fall keine Drift-Probleme hat (Tackmann et al. 2014).

Weitere Studien (z.B. Carraz et al. 2015) zeigen das Potential dieser Methode als Gradiometer für künftige Schwerefeldsatellitenmissionen. So könnte eine Nachfolge-Mission von GOCE dieses Messprinzip realisieren.

## 3. Relativistische Geodäsie

Als weiteres innovatives Konzept in der Geodäsie bieten sich hochpräzise Uhren für die Schwerefeldbestimmung an. Die Gangraten der Uhren hängen gemäß der Einsteinschen Theorie vom Gravitationspotential *U* am Ort der Uhr und von deren Geschwindigkeit *v* ab (Einstein 1916, Moyer 1981). Es gilt für das Verhältnis der Eigenzeit $\tau$ der Uhr und der Koordinatenzeit *t* in erster nach-Newtonscher Näherung:

$$\frac{d\tau}{dt} = 1 - \frac{1}{2}\frac{v^2}{c^2} - \frac{U}{c^2}. \qquad (2)$$

$c$ ist die Lichtgeschwindigkeit im Vakuum. Diese Zusammenhänge sind längst bekannt und werden etwa auch in der Auswertung der geodätischen Weltraumverfahren berücksichtigt, bei GNSS sogar hardwaremäßig korrigiert (Ashby 2002). Inzwischen ist die Genauigkeit der Uhren, speziell optischer Uhren, so hoch, dass man grundsätzlich einen Gangratenunterschied der Uhren aufgrund der Änderung des Schwerepotentials bei einer Höhenvariation von wenigen Zentimetern messen kann (Nicholson et al. 2015, Lisdat et al. 2015). Auf diese Möglichkeit hat bereits Bjerhammar (1985) hingewiesen, der auch den Begriff ‚chronometric leveling' für diese Verfahren einführte; siehe auch Petit und Wolf (1997), Mai (2013) sowie Mai und Müller (2013).

Betrachtet man das Eigenzeitverhalten zweier Uhren, die sich auf der Erdoberfläche befinden, und fasst den Anteil aufgrund der Rotationsgeschwindigkeit der Erde am jeweiligen Uhrenstandort und das zugehörige Gravitationspotential zum Schwerepotential $W$ zusammen, erhält man

$$\frac{d\tau_1}{d\tau_2} = 1 - \frac{\Delta W_{12}}{c^2}. \tag{3}$$

$\Delta W_{12} = W_{P1} - W_{P2}$ gibt die Schwerepotentialdifferenz zwischen den beiden Uhrstandorten an. Hilfreich ist es oft, statt der Eigenzeiten die zugehörigen Eigenfrequenzen zu betrachten, die sich in der Praxis einfacher vergleichen lassen. Sie sind indirekt proportional zu den Eigenzeiten:

$$\frac{df_2}{df_1} = 1 - \frac{\Delta W_{12}}{c^2}. \tag{4}$$

Klassischerweise werden in der Geodäsie physikalische Höhen, also solche Höhen die einen Bezug zum Erdschwerefeld haben, aus kombinierten Nivellements und Schweremessungen bestimmt. Zunächst wird aus den nivellierten Höhenunterschieden $\Delta n$ und den Schweremessungen $g$ entlang der Nivellementslinie eine so genannte geopotentielle Kote berechnet:

$$C_P = -\int_0^p g\, dn = \sum_0^P g\Delta n. \tag{5}$$

Die geopotentiellen Kote entspricht der Differenz des Schwerepotentials $W$ zwischen dem Geoid und dem Punkt $P$, also $C_P = W_0 - W_P$. In der Praxis treten mit dieser Methode nun diverse Probleme auf: Die Fehler des Nivellements wachsen mit der Entfernung; während die Genauigkeit über kurze Distanzen noch im Sub-Millimeterbereich liegt, kann sie über 1000 km in den Zentimeter- und Dezimeterbereich ansteigen. Die Nivellements in größeren Gebieten, z.B. den Bundesländern in Deutschland, wurden zu ganz unterschiedlichen Epochen beobachtet; es können Jahrzehnte dazwischen liegen, was zu systematischen Effekten führen kann. Schließlich ist anzumerken, dass das Verfahren sehr zeitaufwendig ist, wenn man große Entfernungen und/oder Höhenunterschiede überbrücken möchte. Dies potenziert sich noch, falls Höhenänderungen erwartet werden und man Wiederholungsmessungen durchführen muss (Feldmann-Westendorff et al. 2016).

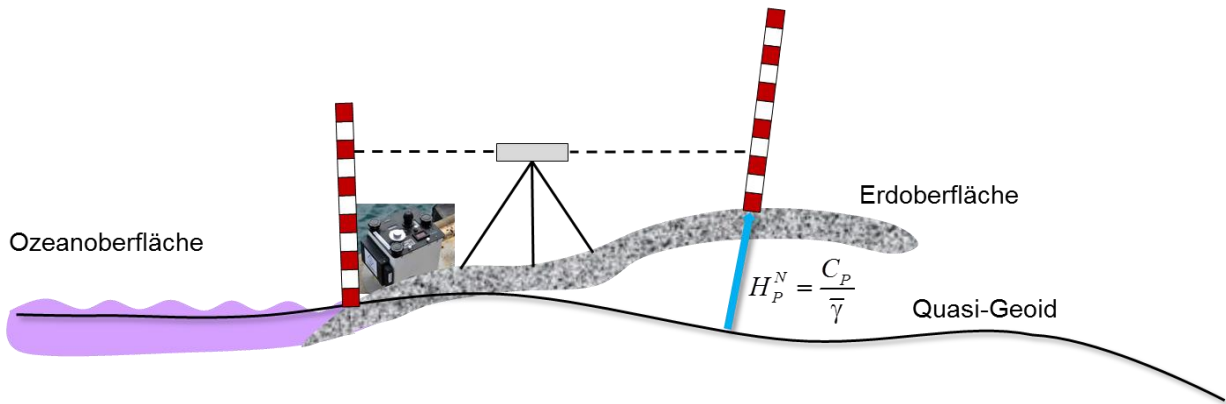

Abb. 6: Definition der Normalhöhe und Messung mittels Nivellement und Gravimetrie. Das Quasi-Geoid ist die Höhenbezugsfläche für Normalhöhen. Es ist eine Näherung an das Geoid, die sich durch die Lösung der Randwertaufgabe nach Molodenskii ergibt (Torge und Müller 2012). Das Quasigeoid stimmt über dem Ozean praktisch mit dem Geoid überein.

Die in Deutschland gebräuchliche Normalhöhe $H^N$ ist für einen Oberflächenpunkt $P$ definiert (Torge und Müller 2012) gemäß

$$H^N = \frac{C_P}{\bar{\gamma}}. \qquad (6)$$

Die geopotentielle Kote $C_P$ wird hier durch die mittlere Normalschwere $\bar{\gamma}$ dividiert, die entlang der Normallotlinie zum Ellipsoid gegeben ist. Meist werden Differenzen zwischen zwei Punkten verwendet, da man bei einem bekannten Höhenpunkt anschließen kann. Dann werden nur noch Differenzen von geopotentiellen Koten $\Delta C_{21} = C_{P2} - C_{P1}$ benötigt, die bis auf das Vorzeichen wiederum identisch mit den Schwerepotentialunterschieden $\Delta W_{21} = -\Delta C_{21}$ sind. Für die Höhendifferenz zwischen zwei Punkten gilt somit ($\Delta \bar{\gamma}_{12} = \bar{\gamma}_{P1} - \bar{\gamma}_{P2}$)

$$\Delta H^N_{21} = H^N_2 - H^N_1 = H^N_1 \frac{\Delta \bar{\gamma}_{12}}{\bar{\gamma}_2} + \frac{\Delta C_{21}}{\bar{\gamma}_2} \qquad (7)$$

bzw. für den ‚Neupunkt' $P_2$ $\qquad (8)$

$$H^N_2 = H^N_1 \frac{\bar{\gamma}_1}{\bar{\gamma}_2} - \frac{\Delta W_{21}}{\bar{\gamma}_2}.$$

Die Potentialunterschiede $\Delta W_{21} = W_{P2} - W_{P1}$ können gemäß Gleichung (4) aus Frequenzunterschieden $\Delta f_{21} = f_{P2} - f_{P1}$, die mit hochpräzisen Uhren gemessen werden, gewonnen werden, und man erhält schließlich

$$H_2^N = H_1^N \frac{\bar{\gamma}_1}{\bar{\gamma}_2} - \frac{c^2}{\bar{\gamma}_2} \frac{\Delta f_{21}}{f_1}. \tag{9}$$

Wenn die relative Frequenz $\Delta f_{21}/f_1$ mit einer Genauigkeit von $10^{-18}$ gemessen wird, entspricht dies einer Höhengenauigkeit von 1 cm. Viele der oben beim Nivellement angesprochenen Fehlerquellen spielen hier keine Rolle mehr. Die Genauigkeit der Schwerepotentialmessung mit Uhren ist unabhängig von der Entfernung. Es wurde außerdem gezeigt, dass die Verbindung der Uhren für kontinentale Entfernungen mit hinreichender Genauigkeit über Glasfaserkabel realisiert werden kann (Droste et al. 2013), dadurch also keine weiteren Fehler auftreten. Weiterhin kann mit Hilfe dieses Messverfahrens die Qualität der gängigen Methoden der Geoid- bzw. Quasi-Geoid-Bestimmung (geometrisch versus gravimetrisch) kontrolliert werden, siehe z.B. Denker et al. (2015).

Bei den Uhrenvergleichen – gerade bei diesem Genauigkeitsniveau – ist aus geodätischer Sicht noch darauf zu achten, jegliche zeitvariablen Anteile, die an den verschiedenen Uhrenstandorten unterschiedlich wirken, konsistent zu korrigieren. Zum Beispiel macht der Effekt durch die festen Erdgezeiten zwischen Braunschwieg und London 8 cm aus, was deutlich über der oben genannten Genauigkeit liegt (Voigt et al. 2016, worin auch weitere Effekte diskutiert werden).

Sollen Uhren während des Transports (mit dem Auto oder Schiff) verglichen werden, was technisch nochmals anspruchsvoller ist als feststehende Uhren, sind weitere relativistische Terme, wie der Sagnac-Effekt, zu berücksichtigen. Denker und Svitlov (2016, priv. Komm.) haben für eine Uhr, die mit dem Auto (durchschnittliche Geschwindigkeit 80 km/h) von Braunschweig nach Paris gebracht wird, einen Effekt von ca. 2 ns ausgerechnet, um den die bewegte Uhr gegenüber einer auf dem Geoid ruhenden Uhr „vorgeht".

In 10 Jahren erwartet man, dass man transportable Uhren hat, mit denen die Messung von Höhenunterschieden mit 1 cm Genauigkeit möglich ist. Weiterhin wird es mobile, optische Uhren geben, die während des Transportes, z.B. auf Schiffen, laufen und kontinuierlich mit einer feststehenden Uhr verglichen werden können. Ebenso wird es Uhren im Weltraum geben, die via Satelliten-Links miteinander verbunden sind. Mit dieser Technologie wird das chronometrische Nivellement als Standard-Technik für die Höhenbestimmung in der Geodäsie etabliert sein, und man wird über neue Konzepte zur Realisierung eines Welthöhensystem basierend auf Frequenzen nachdenken. Darüber hinaus werden sich noch weitere Anwendungsmöglichkeiten entwickeln, etwa Kontrollmessungen von Meeresspiegelvariationen auf Schiffen usw.

## 4. Zusammenfassung und Ausblick

Die genannten Entwicklungen zur atominterferometrischen Schweremessung und zur Erfassung von Schwerepotentialdifferenzen mit präzisen optischen Uhren haben einen Denkprozess in der Geodäsie gestartet. Mittels fallender Atome oder aus der variierenden Tickrate von Uhren die Schwerkraft auszumessen, was zunächst utopisch klingt, ist Realität geworden. Durch die gezielte Nutzung dieser neuen Technologien lassen sich neue Messkonzepte etablieren, die deren Vorteile (wie gleichzeitige Absolut- und Relativgravimetrie oder direkte Messung von Potentialdifferenzen) ausnutzen. Nur die enge interdisziplinäre Kooperation zwischen Physik und Geodäsie erlaubt eine optimale Gestaltung solcher neuer verbesserter Messmethoden für die Schwerefeldbestimmung. Durch die konsequente Anwendung der Quantenphysik und der Relativitätstheorie werden neue Werkzeuge für die Erdbeobachtung entstehen, die für nahezu alle Geo-Disziplinen vorteilhaft sein werden. Der Geodäsie kommt hier eine entscheidende Rolle zu, um dieses Potential wirklich auszuschöpfen.

Grundlegende Arbeiten in diesem Sinne laufen im DFG-Sonderforschungsbereich SFB 1128 ‚Relativistic Geodesy and Gravimetry with Quantum Sensors (geo-Q)' an der Leibniz Universität Hannover. Die erste Phase des SFB 1128, dessen Sprecher Jakob Flury vom Institut für Erdmessung ist, hat im Oktober 2014 begonnen, weitergehende Informationen finden sich auf www.geoq.uni-hannover.de.

**Literatur**

Ashby, N. (2002): Relativity and the Global Positioning System. Physics Today 55 (5), p. 41-47. doi: 10.1063/1.1485583

Bjerhammar, A. (1985): On a relativistic geodesy. Bulletin géodésique 59, p. 207–220

Carraz, O., Siemes, C., Massotti, L., Haagmans, R., Silvestrin, P. (2015): A Spaceborne Gradiometer Concept Based on Cold ATom Interferometers for Measuring Earth's Gravity Field. Microgravity Sci. Technol. 26, p. 139-145, doi: 10.1007/s12217-014-9385-x

Denker, H., Timmen, L., Voigt, C. (2015): Gravity field modelling for optical clock comparisons, 26th General Assembly of International Union of Geodesy and Geophysics (IUGG), June 22 - July 2, Prague, Czech Republic

Droste, S., Ozimek, F, Udem, Th., Predehl, K., Hänsch, T.W., Schnatz, H., Grosche, G., Holzwarth, R. (2013): Optical-Frequency Transfer over a Single-Span 1840 km Fiber Link. Phys. Rev. Lett. 111, 110801, doi: 10.1103/PhysRevLett.111.110801

Eicker, A., Forootan, E., Springer, A. Longuevergne, L., Kusche, J. (2016): Does GRACE see the terrestrial water cycle 'intensifying'? In: Journal of Geophysical Research – Atmosphere 121(2), p. 733–745

Einstein, A. (1916): Die Grundlage der allgemeinen Relativitätstheorie. Annalen der Physik, Verlag von Johann Ambrosius Barth, Leipzig, Vol. 49, p. 769-821

Elsaka, B., Raimondo, J., Brieden, P., Reubelt, T., Kusche, J., Flechtner, F. Iran Pour, S., Sneeuw, N., Müller, J. (2014): Comparing seven candidate mission configurations for temporal gravity field retrieval through full-scale numerical simulation. Journal of Geodesy, Vol. 88, No. 1, p.31- 43, doi: 10.1007/s00190-013-0665-9

Feldmann-Westendorff , U., Liebsch, G., Sacher, M., Müller, J., Jahn, C. Klein, W., Liebig, A., Westphal, K. (2016): Das Projekt zur Erneuerung des DHHN: Ein Meilenstein zur Realisierung des integrierten Raumbezugs in Deutschland. zfv 5/2016, August 2016 accepted

Freier, C., Hauth, M., Schkolnik, V., Leykauf, B., Schilling, M., Wziontek, H., Scherneck, H., Müller, J., and Peters, A. (2016): Mobile quantum gravity sensor with unprecedented stability. Journal of Physics: Conference Series 723, p. 012050, doi: 10.1088/1742-6596/723/1/012050

Gitlein, O. (2009): Absolutgravimetrische Bestimmung der Fennoskandischen Landhebung mit dem FG5-220., Wissenschaftliche Arbeiten der Fachrichtung Geodäsie und Geoinformatik der Universität Hannover, Nr. 281

Hauth, M, Freier, C, Schkolnik, V, Senger, A, Schmidt, M, and Peters, A (2013): First gravity measurements using the mobile atom interferometer GAIN, Applied Physics B 113, p. 49–55, doi: 10.1007/s00340-013-5413-6

Lisdat, C., Grosche, G., Quintin, N., Shi, C., Raupach, CS.M.F.., Grebing, C., Nicolodi, D., Stefani, F., Al-Masoudi, A., Dörscher, S., Häfner, S., Robyr, J.L., Chiodo, N., Bilicki, S., Bookjans, E., Koczwara, A., Koke, S., Kuhl, A., Wiotte, F., Meynadier, F., Camisard, E., Abgrall, M. Lours, M., Legero, T., Schnatz, H., Sterr, U., Denker, H., Chardonnet, C., Le Coq, Y., Santarelli, G., Amy-Klein, A., Le Targat, R., Lodewyck, J., Lopez, O. Pottie, P.-E. (2015): A clock network for geodesy and fundamental science. arXiv: 1511.07735

Mai, E. (2013): Time, Atomic Clocks, and Relativistic Geodesy. DGK, Reihe A, Nr. 124, Beck, München, http://dgk.badw.de/fileadmin/docs/a-124.pdf

Mai, E., Müller, J. (2013): General Remarks on the Potential Use of Atomic Clocks in Relativistic Geodesy. zfv, 4/2013, 138. Jahrgang, p. 257-266

Moyer, T.D. (1981): Transformation from proper time on Earth to coordinate time in solar system barycentric space-time frame of reference. Celestial Mechanics 23, p. 33-56, doi: 10.1007/BF01228543

Müller, J., Naeimi, M., Gitlein, O., Timmen, L., Denker, H. (2012): A land uplift model in Fennoscandia combining GRACE and absolute gravimetry data. Physics and Chemistry of the Earth, 53-54, p. 54-60, doi: 10.1016/j.pce.2010.12.006

NGGM-D Team with contributions by Baldesarra, M., Brieden, P., Danzmann, K., Daras, I., Doll, B., Feili, D., Flechtner, F., Flury, J., Gruber, T., Heinzel, G., Iran Pour, S., Kusche, J., Langemann, M., Löcher, A., Müller, J., Müller, V., Murböck, M., Naeimi, M., Pail, R., Raimondo, J. C., Reiche, J., Reubelt, T., Sheard, B., Sneeuw, N., Wang, X. (2014): e$^2$.motion – Earth System Mass Transport Mission (Square) – Concept for a Next Generation Gravity Field Mission – Final Report of Project "Satellite Gravimetry of the Next Generation (NGGM-D)", DGK Reihe B, Nr. 318, München, ISBN 978-3-7696-8597-8, www.dgk.badw.de/fileadmin/docs/b-318.pdf

Nicholson, T.L., Campbell, S.L., Hutson, R.B., Marti, G.E., Bloom, B.J., McNally, R.L., Zhang, W., Barrett, M.D., Safronova, M.S., Strouse, G.F., Tew, W.L., Ye, J. (2015): Systematic evaluation of an atomic clock at $2 \times 10^{-18}$ total uncertainty. Nature Communications 6, Article number 6896, doi: 10.1038/ncomms7896

Pail, R. and IUGG Writing Team (2015): Observing Mass Transport to Understand Global Change and Benefit Society: Science and User Needs, An international multi-disciplinary initiative for IUGG; in: Pail, R. (eds.) Deutsche Geodätische Kommission der Bayerischen Akademie der Wissenschaften, Reihe B, Angewandte Geodäsie, Vol. 2015, Heft 320, Verlag der Bayerischen Akademie der Wissenschaften in Kommission beim Verlag C.H. Beck, ISBN (Print) 978-3-7696-8599-2, ISSN 0065-5317

Petit, G., Wolf, P. (1997): Computation of the relativistic rate shift of a frequency standard. IEEE Trans. IM 46(2), p. 201-204

Reubelt, T., Sneeuw, N., Iran Pour, S., Hirth, M., Fichter, W., Müller, J., Brieden, P., Flechtner, F., Raimondo, J.C., Kusche, J., Elsaka, B., Gruber, Th., Pail, R., Murböck, M., Doll, B., Sand, R., Wang, X., Klein, V., Lezius, M., Danzmann, K.; Heinzel, G., Sheard, B., Rasel, E., Gilowski, M., Schubert, C., Schäfer, W., Rathke, A., Dittus, H., Pelivan, I. (2014): Future Gravity Field Satellite Missions. In: Flechtner, F., Sneeuw, N., Schuh, W.D. (eds.) Observation of the System Earth from Space - CHAMP, GRACE, GOCE and future missions. GEOTECHNOLOGIEN Science Report No. 20, p. 165-230, Springer, ISBN (Print) 978-3-642-32134-4, ISBN (Online) 978-3-642-32135-1, ISSN 2190-1635, ISSN (Online) 2190-1643

Rudolph, J., Herr, W., Grzeschik, C., Sternke, T., Grote, A., Popp, M., Becker, D., Müntinga, H., Ahlers, H., Peters, A., Lämmerzahl, C., Sengstock, K., Gaaloul, N., Ertmer, W., Rasel E.M. (2015): A high-flux BEC source for mobile atom interferometers. New J. Phys. 17, p. 065001

Rummel, R., Yi, W., Stummer, C. (2011): GOCE gravitational gradiometry. Journal of Geodesy 85, p. 777-790, doi: 10.1007/s00190-011-0500-0

Schilling, M., Müller, J., Timmen, L. (2012): Einsatz der Atominterferometrie in der Geodäsie. zfv, 3/2012, 137. Jahrgang, p. 185-194

Seitz, F., Müller, J. (2016): Erdrotation. Buchkapitel im „Handbuch der Geodäsie", Band „Erdmessung und Satellitengeodäsie" (Hrsg. R. Rummel), Springer, Berlin, S. 1-29, doi: 10.1007/978-3-662-46900-2_12-2

Sheard, B., Heinzel, G., Danzmann, K., Shaddock, D., Klipstein, W., Folkner, W. (2012): Intersatellite laser ranging instrument for the GRACE follow-on mission. J. Geod. 86(12), p. 1083-1095

Steffen, H., Gitlein, O., Denker, H., Müller, J., Timmen, L. (2009): Present rate of uplift in Fennoscandia from GRACE and absolute gravimetry. Tectonophysics 474, p. 69-77, doi: 10.1016/j.tecto.2009.01.012

Tackmann, G., Berg, P., Abend, S., Schubert, C., Ertmer, W., Rasel, E.M. (2014): Large-area Sagnac atom interferometer with robust phase read out. Comptes Rendus Physique 15, p. 884-897, doi: 10.1016/j.crhy.2014.10.001

Timmen, L., Gitlein, O., Klemann, V., Wolf, D. (2011): Observing gravity change in the Fennoscandian uplift area with the Hanover absolute gravimeter, In: Deformation and gravity change: indicators of isostasy, tectonics, volcanism and climate change

Vol. III, Pure Appl. Geophys. (PAGEOPH), Springer Basel AG, doi: 10.1007/s00024-011-0397-9

Torge, W., Müller, J. (2012): Geodesy. 4th edition, de Gruyter, Berlin/Boston

Vestøl, O., Ågren, J., Steffen, H., Kierulf, H., Lidberg, M., Oja, T., Rüdja, A., Kall, T., Saaranen, V., Engsager, K., Jepsen, C., Liepins, I., Paršeliūnas, E., Tarasov L. (2016): NKG2016LU, an improved postglacial land uplift model over the Nordic-Baltic region. Presentation at NKG WG meeting in Tallinn, 12-18 March 2016

Voigt, C., Denker, H., Timmen, L. (2016): Time-variable gravity potential components for optical clock comparisons and the definition of international time scales. Metrologia, accepted 29 July 2016